# A Cryptographic Escrow for Treaty Declarations and Step-by-Step Verification

*Sébastien Philippe*[*,†], *Alexander Glaser*[**], *and Edward W. Felten*[***]

[*]Belfer Center for Science and International Affairs, John F. Kennedy School of Government, Harvard University, Cambridge, MA 02138, USA

[**]Program on Science and Global Security, Princeton University, Princeton, NJ 08542, USA

[***]Center for Information Technology Policy, Princeton University, Princeton, NJ 08544, USA

[†]Corresponding author: sebastien_philippe@hks.harvard.edu

**Abstract.** The verification of arms-control and disarmament agreements requires states to provide declarations, including information on sensitive military sites and assets. There are important cases, however, where negotiations of these agreements are impeded because states are reluctant to provide any such data, because of concerns about prematurely handing over militarily significant information. To address this challenge, we present a cryptographic escrow that allows a state to make a complete declaration of sites and assets at the outset and commit to its content, but only reveal the sensitive information therein sequentially. Combined with an inspection regime, our escrow allows for step-by-step verification of the correctness and completeness of the initial declaration so that the information release and inspections keep pace with parallel diplomatic and political processes. We apply this approach to the possible denuclearization of North Korea. Such approach can be applied, however, to any agreement requiring the sharing of sensitive information.

## Introduction

Ever since the Strategic Arms Limitation Talks between the United States and the Soviet Union, nuclear arms-control treaties have involved transparency measures and the exchange of information[1]. Negotiating deep cuts in U.S. and Russian nuclear arsenals would require unprecedented disclosures, however. In 1997, a National Academy of Sciences study proposed these transparency measures could include[2]: "the current location, type, and status of all nuclear explosive devices and the history of every nuclear explosive device manufactured, including the dates of assembly and dismantling or destruction in explosive tests; a description of facilities at which nuclear explosives have been designed, assembled, tested, stored, deployed, maintained, and dismantled, and which produced or fabricated key weapon components and nuclear materials; and the relevant operating records of these facilities." Such disclosures are difficult to undertake, because they could provide an adversary with military significant information at an early stage. In a 2005 study, the National Academy of Sciences' Committee on International Security & Arms Control suggested that cryptography could help address this problem[3].

A similar challenge is arising today in the context of U.S. – Democratic People's Republic of Korea (DPRK) talks on the denuclearization of the Korean Peninsula. This is because denuclearization shares many of the problems associated with deep nuclear reductions in a very acute way. As part of an agreement, the DPRK would most likely be required to provide data and disclose activities related to its nuclear and ballistic-missile programs, as well as submit to observation and onsite inspections by the international community[4–6]. Prior verification plans proposed by the United States[7], asked the DPRK to make substantial and detailed baseline declarations including: the current location, type, and status of all nuclear weapons and associated components; a description of facilities at which nuclear materials and weapons have been produced, designed, assembled, tested, stored, and deployed; and data on the quantities and characteristics of declared nuclear material. From the DPRK point of view, agreeing to such demands may be too risky: it would provide the United States with a potentially comprehensive map of its military and nuclear weapons-related assets at a very early stage in the diplomatic process, which could become an important security threat if negotiations were to collapse. But given the strong U.S. public commitment to verifiable denuclearization, it is difficult to conceive a successful diplomatic outcome in which the DPRK would not provide any kind of useful declaration[8,9].

Here, we address this negotiation challenge by presenting an approach to declarations that provides a secure information-sharing mechanism for a state to sequentially reveal relevant sensitive information to another state while requiring the country making declarations to commit to the correctness and completeness of their initial declaration at the outset, potentially even before negotiations start. This cryptographic escrow scheme then enables the release of partial information for verification at later stages, as opposed to engaging in the full disclosure of all data at once (Fig. 1). This allows data exchanges to keep pace with confidence-building measures.

Our escrow leverages cryptographic primitives, in particular commitment schemes[10]. Such schemes allow a party to commit to a particular piece of information, or value, while keeping it hidden from others. The committing party can release the value at a later stage while ensuring other parties it was not altered.

While verifying the denuclearization of North Korea is a particularly relevant application for our approach, similar escrow scheme could be used in other international agreements including the exchange of secure declaration as part of future U.S. – Russia arms-reduction efforts[3], or the declaration of sensitive information (e.g. identification and location of pollution emitters) in environmental agreements[11,12].

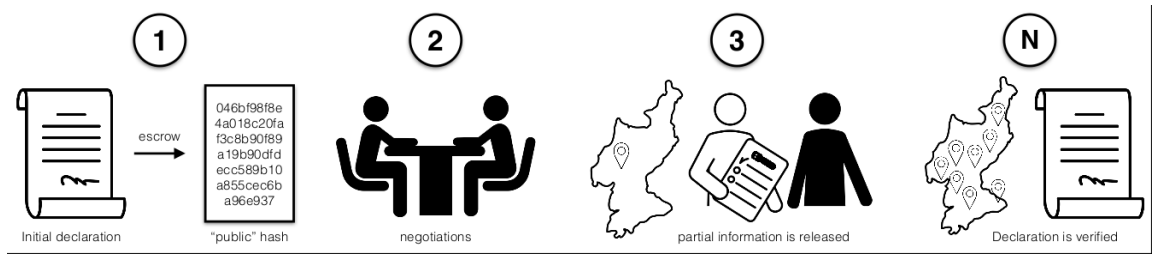

**Figure 1. Using a cryptographic escrow in an inspection regime.** (1) A detailed initial declaration is produced by the inspected party and placed in an escrow. A cryptographic commitment to this declaration is made public. (2) The negotiations are ongoing. The escrow is built such that is possible to reveal only partial information at a time. (3) Prior to an on-site inspection, partial information about a site (location, status and items) is revealed to the inspecting party. The inspections eventually confirm the correctness of this information. (N) As negotiations move forward, information is released incrementally until the complete declaration is revealed. Only then the inspecting party has a complete picture of the inspected party assets.

## Escrow Construction

The most basic construction for our escrow could be a cryptographic commitment of the entire declaration. One way to implement a commitment scheme is through the application of cryptographic hash functions. In general, the hash of a message is much shorter than the message itself, and the underlying cryptographic hash function is designed such that it is infeasible to find a valid message for a given hash (assuming the values being hashed are drawn from a random distribution with high entropy) and infeasible to construct two different messages that produce the same hash (a property called collision resistance). In principle, multiple hash functions can be combined, using robust multi-property combiners, so that each of the necessary cryptographic properties holds for the combination if it holds for at least one of the hash functions being combined[13]. This could be used, for example, to allow each party in our scheme to propose a hash function he or she trusts, and to use a combined hash function that has the desired security if either one of the parties' chosen hash functions is secure.

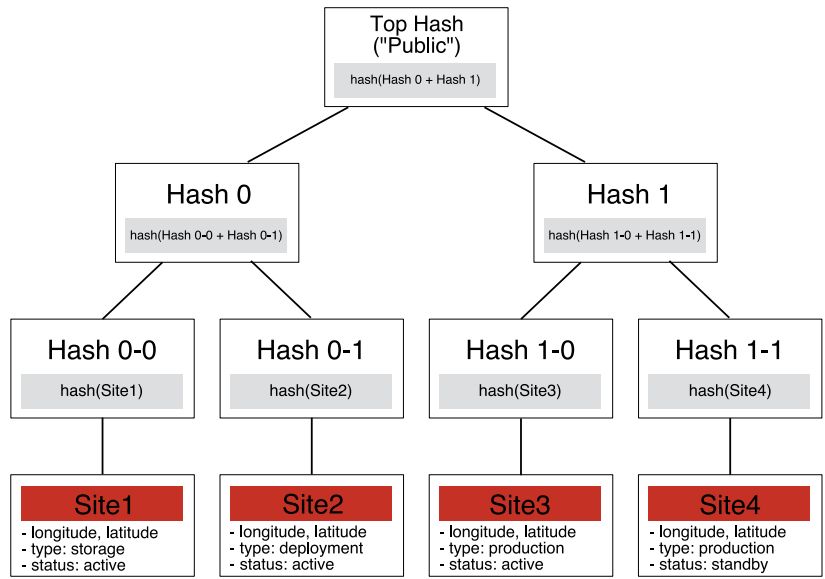

**Figure 2. Sites declaration using a Merkle Tree structure.** Each leaf of the tree contains information on individual sites. The information of each data block is hashed individually, the upper nodes hashes "0" and "1" are obtained by hashing the concatenation of the two lower hashes up to the Top Hash or root of the tree. To later demonstrate that the information of Site 4 was part of the declaration, the prover needs to supply the clear text of Site 4, the hash of Site 4 (Hash 1-1), the hash 1-0, the Hash 1 and the Top Hash. The process does not reveal information about any other sites.

Simply committing to the complete declaration would not provide for any flexibility on how much and what information can be revealed at a time, however. To address this issue, we turn our escrow into a binary Merkle tree[14] (see Fig. 2). The tree is constructed as follows: every leaf (or childless node) can store any string, for example, a cryptographic commitment to a data block with information related to a specific site, including, in the case of North Korea, the denuclearization-relevant items stored at the site. Any non-leaf node must store the value *Hash*(*L*,*R*) whenever its left child stores *L* and right child stores *R*. The root of the tree then represents a commitment to the entire declaration and would be the only piece of information made public at the beginning of the diplomatic process.

Furthermore, we build the tree such that a pair of geographic coordinates in the country corresponds to a unique leaf in the tree. To do so, we superpose a grid over the map of country (see Fig. 3). For North Korea, the grid is limited by latitudes 37.5° N and 43.5° N, and longitudes 124.0° E and 131.0° E. We then construct a local coordinate system (*i*,*j*) with the point (43.5° N, 124.0° E) as the origin. The number of grid points depends on the chosen resolution in latitude and longitude. For a resolution of one minute in both latitude and longitude (corresponding to ~1.15 miles N-S and ~0.89 miles E-W), as called for in existing arms-control agreements requiring the sharing of coordinates information[15], there will be $7*6*60^2$ = 151200 points. Each point is numbered individually using its local coordinates (*i*,*j*). The numbers are converted in base 2 and concatenated to obtain the corresponding binary key *x*. For example, the point of coordinates (7*60, 6*60) on the one-minute resolution grid corresponds to the key *x* = 110100100101101000 of length *l* = 18 bits.

Because the number of grid points is not too large, we opt for a simplified construction in which there is a leaf node for every grid point. In scenarios where the number of grid points would make the tree too large to be practical, other cryptographic data structures[16] could provide the necessary properties without requiring the declaring party to store data for every empty grid point.

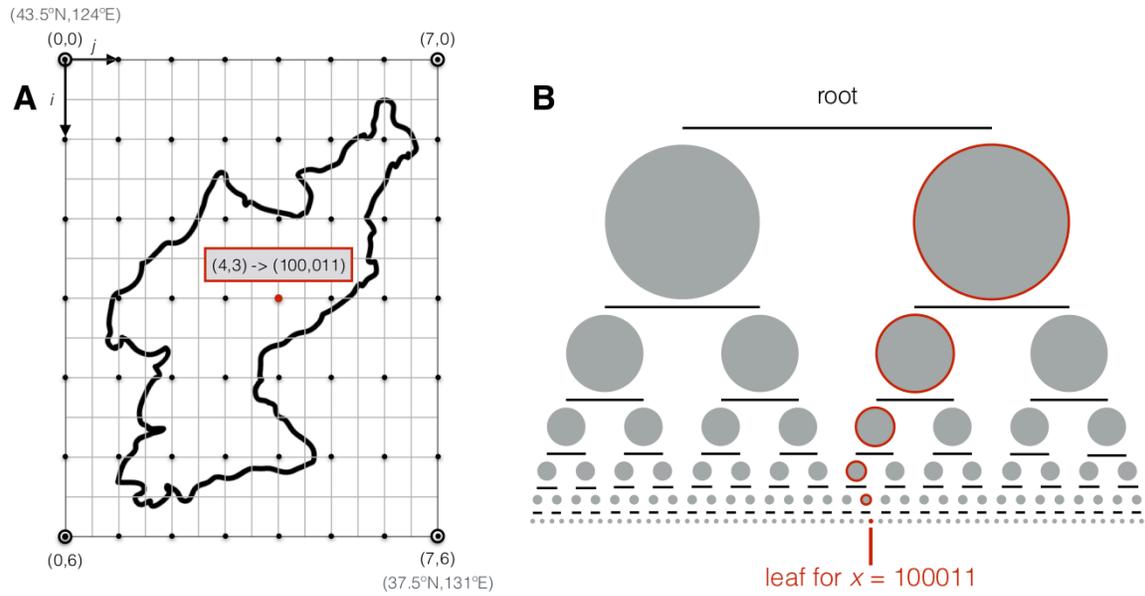

**Figure 3. Mapping a site coordinates to a Merkle tree leaf.** Local coordinates (*i,j*) placed over the DPRK (A) can identify uniquely each location in the country with a given precision. The numbers *i* and *j* are then converted in their binary equivalent and concatenated into a string *x* corresponding to the binary path from the root to the corresponding leaf in our Merkle tree (B). Each leaf of the tree either stores a commitment to information about an existing site or to nothing if no site is present at this location.

In a tree of depth *l*, each leaf is then reached uniquely from the root via the path defined by *x* (see Fig. 3), and contains a commitment on whether or not there is a site at this location and any other relevant information about the site, if it exists. In the case of North Korea, the number of sites to be declared is expected to be ~ 100–200[17], much smaller than the number of leafs, $2l$. This construction allows information about any grid point, or any subset of grid points, to be revealed without conveying information about any other grid points.

## Step-by-Step Verification in a Freeze Scenario

Figure 4 presents a cryptographic commitment (using a hash function) to information about a hypothetical nuclear weapons storage site, which would be stored in the leaf in the tree corresponding to the site location. The commitment is obtained by hashing a message "m_0" containing different pieces: a random number generated by the committing party, another random number provided by an outside party to guarantee freshness of the information[18] (i.e., that the commitment was produced after the outside party's random number was generated), information regarding the entry including for example the type of facility, coordinates, and status, and finally additional information "m_1", which represent commitments to additional data that may be shared at a later stage, for example prior to the inspection of the site in question.

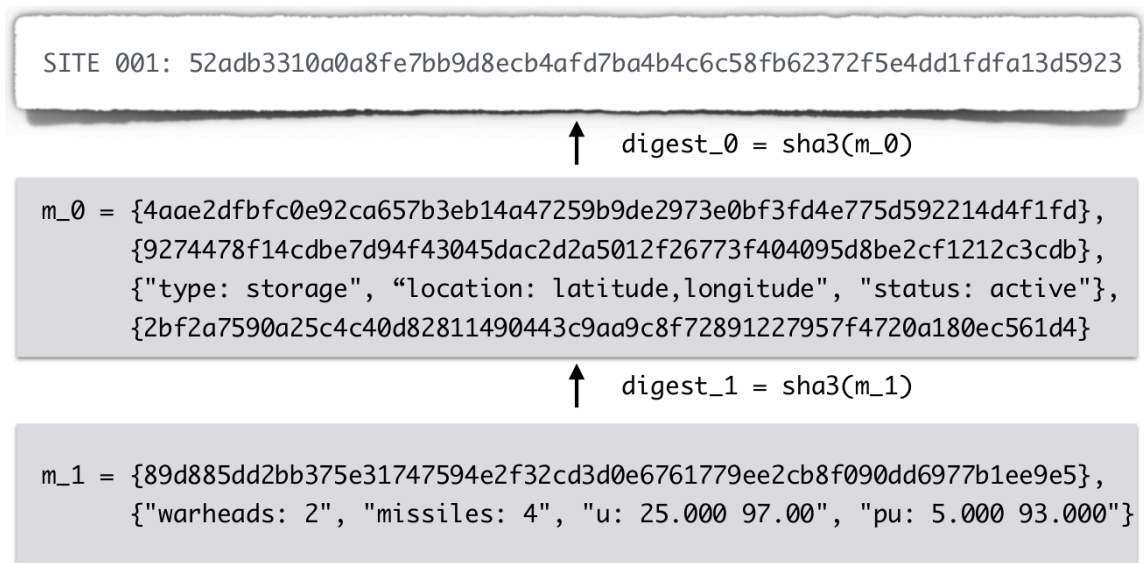

**Figure 4. Example of message digest for a storage site.** Data is encapsulated in multiple levels, which can be revealed at different points in time. The level 0 message contains a random number generated by the host, a random number provided by inspectors to guarantee freshness of the commitment, clear text data and the hash of the level 1 message. The level 1 message contains additional data (here the number of warheads, missiles, amounts of uranium and plutonium, and their isotopics), which may be released at a later stage, for example prior to inspection. The message digests (generated with SHA3-256) are for illustration purpose only.

As a preliminary step in a verified denuclearization process following a phased approach[4], the DPRK could agree on freezing the production of fissile materials and components for weapons as well as on monitored storage of existing weapons. Under this framework, the DPRK would produce a complete escrow of all production, storage, and deployment sites of nuclear weapons, missiles, and associated components. It would then commit to the inventories at each site and agree not to move assets between sites. (Movement patterns between sites could be monitored with satellites.)

To verify correctness of the declaration, the DPRK would invite the United States to perform onsite inspections and verify that the assets and information declared in the escrow are present and valid. During these inspections, accountable items could be tagged with unique identifiers[19], the United States would become more confident that a freeze is indeed in effect and that the rest of the declaration, which has yet to be revealed, is correct.

Confidence from the U.S. point of view would increase if sites could be picked at random[20], although the DPRK may prefer to reveal the location and inventories at each site in the order it decides, for example starting with sites that are already known or considered less sensitive. Because each site can be revealed without compromising others, the pace of inspections can be adapted to the political process, making this approach well suited for an "action for action" negotiating process, where both side would make incremental concessions working towards an ultimate settlement.

Combining the properties of the escrow and the possibility to perform challenge inspections would facilitate the process of establishing completeness of the declaration. If the United States believes it has detected proscribed activities at an undeclared site, it could provide North Korea with the site coordinates, corresponding to a specific key $x$. The DPRK could then prove whether or not it has included this specific site in the escrow. If the site is in the escrow, both parties would wait and plan for a future inspection to confirm the correctness of the declaration. If the site is not in the escrow, a special inspection would have to take place to demonstrate that no proscribed activities are taking place at the site. Given this risk of exposure, it would be in the interest of the DPRK to produce a complete declaration from the very beginning.

## Security of the Escrow

Overall, for the viability of our approach, it is imperative that the message length, the message content, and the implementation of the commitment protocol are robust against all relevant types of cryptographic attacks. When using hash functions, it is important to be mindful of the potential for preimage, and collision attacks[21]. Preimage attacks would allow finding a message corresponding to a given hash produced with a particular hash function. (If the length of a hash is $n$ bits, a brute force attack would requires $\sim 2^{\frac{n}{2}}$ evaluations of the hash function for finding a collision between two messages, and $\sim 2^n$ evaluations for finding preimages and second preimages.). These would compromise the information committed in our escrow. So far, there have been no known successful preimage attacks on NIST recommended hash functions[22,23].

What is more typical for past hash functions, however, has been the discovery of collision attacks, which would challenge the binding property of the commitment scheme[24,25]. A recent practical example, is the discovery of the first collision for SHA-1 using a method to produce two PDF documents producing the same hash[26]. Discovery of a SHA-1 collision was anticipated for many years before it occurred, however. In our case, new collision attacks could affect the security of past declarations if they also allow to conduct secondary preimage attacks. These risks could be mitigated by using hash function combiners, allowing multiple hash functions to be combined in such a way that the combination is collision-free if at least one of the constituent hash functions is collision-free[13]. If doubt should arise about the continued collision-freedom of the hash functions being used, commitments could be re-generated to include a new combination of hash functions, to provide additional insurance against collisions.

## Conclusion

North Korean diplomats could walk to the negotiation table to meet their U.S. counterparts with a 256-bit or 512-bit message on a piece of paper. Using the escrow scheme developed in this paper, this simple message could represent a commitment to a database containing every single bit of information about their nuclear and ballistic missile programs. Doing so would fulfill a U.S. demand to provide a comprehensive declaration of sites and assets. It would also prevent the United States from walking away with this information, a potentially unacceptable security threat for North Korea.

We showed how to combine our escrow with an inspection regime to verify the correctness and completeness of a sites declaration in a step-by-step approach. While not all information is available upfront to inspectors, confidence in the validity of the overall declaration grows with each successful inspection. Our approach also allows the inspected party to commit to additional information documenting weapon design, production records, and movement of assets through the weapons complex.

The approach presented in this paper has the potential to resolve a longstanding diplomatic deadlock: the United States wants a correct and complete declaration from the DPRK, which in return does not want to provide a target list that could enable a preventive military attack. Our proposal resolves this tension by allowing the DPRK to commit to such a declaration, which is gradually revealed as the diplomatic process proceeds. In the longer term, the case of North Korea could serve as an important precedent for using modern cryptographic techniques to support nuclear arms-control and disarmament.

**Acknowledgements.** The authors thank B. Barak, R.J. Goldston, F. von Hippel, and Z. Mian for their comments and feedback.

**Author contributions.** S.P. led the research. A.G. and E.W.F. contributed, along with S.P., to developing the cryptographic escrow for treaty declarations and verification. All authors contributed to the manuscript.

**Competing interests.** The authors declare no competing interests.

**Data availability.** All data are available from the corresponding author upon request.